\documentstyle[11pt]{article}
\newcommand{\be}{\begin{eqnarray}}

\newcommand{\ee}{\end{eqnarray}}

\title {
\begin{flushright}
{\normalsize TPI-MINN-93/56-T \\
  NUC-MINN-93/31-T \\
  UMN-TH-1228/93 \\
  November 1993 \\}
\end{flushright}
\bf
 Recent Progress in Understanding
  Quark and Gluon Distribution Functions for Large Nuclei
  at Small x\footnote{Presented at Hot and Dense Nuclear Matter,
NATO Advanced Study Institute, Bodrum, Turkey, Sep. 26 - Oct. 9, 1993}
}
\author{
  Larry McLerran\\
  {\small\it Theoretical Physics Institute,} \\
  {\small\it School of Physics and Astronomy,} \\
  {\small\it University of Minnesota ,}\\
  {\small\it Minneapolis, MN 55445}
}
\date{}

\begin{document}

\maketitle
\begin{center}
{\bf Abstract} \\
\end{center}

I discuss the problem of computing the structure functions for very heavy
nuclei at small Bjorken x.  The approximations used in this description
are physically motivated, and recent computations reviewed.

\vfill \eject
\section{INTRODUCTION}

The theoretical question which will be addressed in this talk is
the computation of quark and gluon distribution functions at small
Bjorken x.   The reason why such computations may be possible
is because at small values of Bjorken x, which correspond to
central values of rapidity, the density of partons per unit area
satisfies
\begin{equation}
   {1 \over {\pi R^2}} {{dN} \over {dy}} >> \Lambda^2_{QCD}
\end{equation}
This density of partons per unit area is the only local parameter
with dimensions of an energy squared which we can construct, and
the coupling therefore should be evaluated at this scale.  Defining
\begin{equation}
	\Lambda^2 = {1 \over {\pi R^2}} {{dN} \over {dy}}
\end{equation}
the strong coupling constant at this scale must be
\begin{equation}
	\alpha_s(\Lambda ) << 1
\end{equation}
so that it may be possible to formulate the problems in weak
coupling.

There are several problems which might be solved if one can
compute the quark and gluon distribution functions at small
Bjorken x.  At small x, it is expected that the gluon distribution
function for a single proton behaves as \cite{lipatov}
\begin{equation}
    {{dN} \over {dx}} \sim {1 \over {x^{1+C\alpha_s}}}
\end{equation}
This implies that the distribution of partons for a proton diverges as
$x \rightarrow 0$.  If this is the case, then this problem should be
possible to analyze using weak coupling methods.
The distribution functions at small $x $ have recently
been measured at HERA, and do seem to be singular at small $x$.

In addition to
understanding the Lipatov enhancement, it would also be useful
to compute quantities such as the ratio of sea quark to gluon
distributions.  For example, if the typical energy scale has been increased
due to a high density of partons, we should expect that heavier quarks
will become of increasing importance.  Charm quark production might for
example become substantial.

Another interesting physical problem is deep inelastic scattering
and di-lepton production using nuclei with $A >> 1$.  If the value of
Bjorken $x << A^{-1/3}$, then the nucleus as seen by a parton moving
with that value of $x$ is Lorentz contracted to a scale size which
is much smaller than the wavelength of the parton in a frame
comoving with its longitudinal momentum.  It is expected that
in this kinematic region, there should be non-trivial effects which
might screen the effects of the valence nuclear matter distribution.
On the other hand, we expect that the distribution functions for
partons should become large for large nuclei, and if the effects
due to screening can be ignored (as we will see they can be when
we do the computation) then
\begin{equation}
	{1 \over {\pi  R^2}} {{dN} \over {dy}} \sim A^{1/3} >>
\Lambda^2_{QCD}
\end{equation}
as $A \rightarrow \infty$.  In addition to the questions about the
Lipatov enhancement and the ratio of sea quarks to glue,  one can
also ask about the $A$ dependence of the distribution functions.

Finally, there is the problem of determining the initial conditions
for quarks and gluons in heavy ion collisions.  If one can understand
the distribution functions, then these may provide information about
the boundary conditions for the evolution of the matter into a
quark-gluon plasma.  Recall that since the density of partons per unit area
is the only scale in the problem, and it goes like $A^{1/3}$, the
energy density will have to scale as $E/V \sim A^{2/3}$, where this
is the energy density scale at a time which corresponds to the
dimensional scale constructed from the density of partons per unit
area.  In order for these considerations to be valid, the Lorentz
contracted size of the nucleus must be smaller than the size scale
constructed from the density of partons per unit area.  This is
$E_{CM} / A >> \sqrt{A}$ which for large nuclei requires that
$E_{CM}/A >> 50~ GeV$ which is within the range accessible at RHIC.

In what follows,  I will describe how to compute the quark and gluon
distribution functions for very large nuclei.  We have not yet
succeeded in being able to compute for a single hadron.  It may also
be true that our weak coupling analysis may be marginal at best for
realistic values of $A$, and therefore our results may only  be useful
as a theoretical laboratory which will give us some insight into
the structure which we might expect for realistic nuclei.  If it is
true that weak coupling techniques cannot be used for the earliest
stages of a  heavy ion collision, then it also will imply that such
techniques are probably never applicable in the subsequent evolution
of the matter produced in such a collision.

\section{Summary of Results}

Before going into a discussion of how to analyze the problem
of computing distribution functions for very large nuclei, I will
first summarize our results to date.\cite{mclerran}  First I define
\begin{equation}
	\mu^2 = {4 \over 3} {N_Q^{valence} \over {\pi R^2}} \sim 1.1
A^{1/3}~Fm^{-2}
\end{equation}
which is the the density of valence quark color charge squared per unit area.
I will show that there is a many body theory which describes the
quark and gluon distribution functions so long as we restrict our
attention to parton transverse momenta which satisfy
\begin{equation}
	q_T^2 << \mu^2
\end{equation}
and we require that we are at small values of $x$
\begin{equation}
	x << A^{-1/3}
\end{equation}
There are two expansion parameters for this many body theory
$\alpha_s(\mu)$ and $\alpha_s(\mu ) \mu /q_T$

In the range $\alpha_s(\mu )  \mu << q_T << \mu $ the gluon
distribution function may be evaluated to lowest order in $\alpha_s
(\mu )$ as
\begin{equation}
	{1 \over {\pi R^2}}  {{dN} \over {dx d^2q_T}} =
{{\alpha_s (N_c^2-1)} \over \pi^2} {{\mu^2} \over {xq_T^2}}
\end{equation}
This is just the Weizsacker-Williams distribution function for
a Lorentz boosted distribution of Coulombic gluons, scaled by the
factor $\mu^2$ which is the average value of the charge squared
per unit area.

The result above is however only a formal result if it is true that
there is a Lipatov enhancement.  The Lipatov enhancement is of the
form $x^{-C\alpha_s}$   If we formally expand this as a series in
$\alpha_s$, we find the expansion parameter is $\alpha_s (\mu )
ln(1/x)$.  Since we have assumed that $x << A^{-1/3}$ and
$\mu^2 \sim A^{1/3}$, this expansion is not well behaved.  To
correct for this behavior if it occurs, it will be necessary to go
beyond the naive weak coupling expansion.  It should however still be
true that weak coupling methods are still applicable.

If we sum to all orders in $\alpha_s \mu$ but to first order in
$\alpha_s$, we have shown that the gluon distribution function is of
the form
\begin{equation}
	{1 \over {\pi R^2}} {{dN} \over {dxd^2q_T}}
 = {{N_c^2 - 1} \over \pi^2}~ {1 \over x}~ {1 \over \alpha_s}
H(q_T^2/\alpha_s^2\mu^2 )
\end{equation}
where the function $lim_{y \rightarrow \infty} H(y) \rightarrow
1/y$  The function $H$ is a correlation function for an
ultraviolet finite two dimensional Euclidean quantum field theory.
The strong coupling limit of this field theory is the small $q_T$
limit.  In this limit, we expect disorder since the theory is strongly
coupled.  This implies exponentially falling correlations in
coordinate space which implies that $H(0)$ should be finite.

\section{Setting the Problem Up}

To begin computing the distribution functions, we will assume we
are working in a frame where the nucleus is moving close to the
speed of light.  In this frame the nucleus appears as a Lorentz
contracted pancake.  The value of Bjorken $x$ in this frame is
essentially the ratio of longitudinal momentum of the parton to that
of the projectile nucleus per nucleon.

Since the variation in the nuclear valence quark distribution
function with respect to transverse coordinate is
\begin{equation}
	{1 \over N} {{dN} \over {dr}} \sim {1 \over R_{nuc}}
\end{equation}
and since the average scale of transverse momenta is
$q_T >> 1/R_{nuc}$, we see that locally, the variation in the
transverse nuclear matter distribution can be ignored.  Therefore to
compute the local properties of the gluon distributions function as a
function of transverse coordinate,we need only consider a problem
where the transverse matter distribution is assumed to be uniform
and of infinite extent in the transverse direction.  The problem is
therefore of a sheet of valence quarks uniformly distributed on a
thin sheet of infinite extent in the transverse direction.
We will take the density of valence quarks to be $N_{quark}/\pi R^2
\sim A^{1/3}$.

The natural variable with which to analyze the dynamics are light
cone variables,
\begin{equation}
	a^\pm = {{a^0 \pm a^z} \over \sqrt{2}}
\end{equation}
If we let $k^+$ be a parton light cone momentum and $p^+$ that of a
projectile nucleon, then $x = k^+/p^+$  In these variables, instead of
constructing eigenstates of the Hamiltonian, it is simpler to
construct the eigenstates of the  generator of $x^+$
transformations,
\begin{equation}
	P^- = {1 \over \sqrt{2}} (H-P^z)
\end{equation}

Finally, when constructing the light cone Hamiltonian and action for
the theory in terms of these variables, it is simplest to work in
light cone gauge,
\begin{equation}
	A_- = -A^+ = 0
\end{equation}
In this gauge, the light cone Hamiltonian has the form
\begin{eqnarray}
	P^- & = & \int d^3x  ~ {1 \over 4} F_T^2  + {1 \over 2}
(\rho_F + D_T \cdot E_T ) {1 \over P^{+2}} (\rho_F + D_T \cdot E_T )
\nonumber \\
 & &
+ {1 \over 2} \psi^\dagger (M-\gamma \cdot P_T ) {1 \over P^+}
(M + \gamma \cdot P_T) \psi
\end{eqnarray}
In this equation,
\begin{equation}
	P^\mu = {1 \over i} \partial^\mu  -g \tau \cdot A^\mu
\end{equation}
and $E_k = -\partial_- A_k$.  The Lagrangean is generated in the
usual
way by adding in $-i \psi^\dagger \partial_+ \psi + E_k \partial_+
A_k$
The fermion charge density is
\begin{equation}
	\rho_F^a =  \overline \psi \gamma^+ \tau^a \psi
\end{equation}
The quantized fields are
\begin{eqnarray}
 \psi_\alpha (x) & = & \int_{k^+ > 0} {{d^3k} \over {(2\pi)^3}}~
(b_\alpha (k) e^{ikx} + d^\dagger_\alpha (k) e^{-ikx} ) \nonumber \\
 A_i^a (x) & = & \int_{k^+ > 0} {{d^3k} \over {(2\pi )^3 \sqrt{2k^+} }}~
(a_i^a (k) e^{ikx} + a_i^{a\dagger} (k) e^{-ikx} )
\end{eqnarray}

The last ingredient we need to set the problem up is how to describe
the valence quarks.  These quarks are traveling close to the speed
of light.  For small values of $\alpha_s$, these quarks occasionally
emit a small $x$ gluon.  This emission does not change the path of
the valence quark.  It should therefore be a good approximation to
treat the trajectories for the valence quarks as straight line
propagation at light velocity, that is
\begin{equation}
	J^+_a = \rho_a (x^+,x_T ) \delta (x^-)
\end{equation}
Unlike the case for QED, in general in QCD the charge density will
have to depend on the time $x^+$ since the extended current
conservation condition requires that
\begin{equation}
	\partial_+ Q^a + f^{abc} A_b Q_c = 0
\end{equation}
This forces the charge to rotate as
\begin{equation}
	\tau \cdot Q (x^+) = U(x^+) \tau \cdot Q(0) U^\dagger(x^+)
\end{equation}
where
\begin{equation}
	U(x^+) = T exp\left( \int_0^{x^+}~dy^{ +}~ \tau \cdot A_+ (y^+) \right)
\end{equation}

To summarize, we have shown the problem which must be solved is
to compute the ground state expectation values for a system with
the valence quarks traveling with light velocity localized along an
infinite sheet in the transverse space.  The density of the valence
quarks is uniform.  This problem is well posed since the
constraint on the valence quarks is equivalent to specifying the
space-time coordinates of the electromagnetic charge and baryon
number.  These operators commute with the QCD Hamiltonian.

\section{The Example of QED}

In  QED,  the problem we want to solve is the photon structure
function generated by a fast moving electron.  We will ignore pair
production of electron-positron pairs.  The source for the electron is
$x^+$ independent and is
\begin{equation}
	\rho_e = e \delta (x^-) \delta^{(2)} (x_T)
\end{equation}
The light cone Hamiltonian is
\begin{equation}
	P^- = \int d^3x~ \left( {1 \over 4} F_T^2 +
{1 \over 2} (\rho_e + \nabla_T \cdot E_T) {1 \over {P^{+2}}}
(\rho_e + \nabla_T \cdot E_T ) \right)
\end{equation}

For this Hamiltonian, the ground state is a coherent state
\begin{equation}
	\mid \Psi > = C exp \left( i \int d^3x A^{op} (x) E^ {cl} (x)
\right) \mid 0 >
\end{equation}
Letting $P^-$ operate on this state, we see that the ground state has
$P^- = 0$ so that the classical field is purely longitudinal
and
\begin{eqnarray}
	B_T & = & 0 \nonumber \\
         \nabla _T \cdot E_T & = & -e \rho_e
\end{eqnarray}
The solution for these equations are that
\begin{equation}
	\vec{A_T} = e~ {1 \over k^+} ~{\vec{k}_T \over {k_T^2}}
\end{equation}
In space-time, the vector potential is $\theta(x^-) \vec{\nabla }
\lambda $ which is for $x^- < 0$ vanishing and a pure gauge for
$x^- > 0$

The field above is precisely the Weizsacker-Williams field for the
Lorentz boosted Coulomb field.  The distribution function for the
photons can be computed and is
\begin{eqnarray}
        F(k^+,k_T ) & = & {1 \over {(2\pi )^3}}~
<a^\dagger (k) a(k) > \nonumber \\
              & = & {{2e^2} \over {(2\pi )^3}} ~{ 1 \over {k^+ k_T^2}}
\end{eqnarray}
or
\begin{equation}
	F(x,k_t) = {\alpha \over \pi^2 }~ { 1 \over {xk_T^2}}
\end{equation}

\section{The Distribution Functions for QCD}

For QCD, we have not been successful in constructing the ground
state wavefunction in the presence of the external source
corresponding to the valence quarks.  We have however concentrated
on computing ground state expectation values.  To do this, we first
note that
\begin{equation}
	Z = lim_{T \rightarrow \infty}~ \sum_N <N\mid e^{iTP^-} \mid N>
\end{equation}
will  project onto the ground state.  The sum over $N$ here includes
a sum over the color labels of the external source of color charge
generated by the valence quarks.

To do the sum over N, we break our transverse space into a grid of
squares with size $d^2x >> \pi R^2/N_{quark} \sim A^{-1/3} Fm^2$
Our approximation will therefore only be good when we look at
transverse momentum scales where $q_t^2 << \mu^2$.  In these
limits, the number of valence quarks in each square is much larger
than 1.   If this is the case, then typically the charge in each
square will be much larger than 1.  If $Q$ is this charge, then $Q^2 >>
Q \sim [Q,Q]$ so that the charge may be treated classically.

If the total charge of interest is also much less than the maximum
possible charge in the square, then the density of states for charge
$Q$ is $e^{-Q^2/2\mu^2}$.  Summing over the states in the definition
of $Z$ is therefore equivalent to inserting into the path integral the
integration
\begin{equation}
	exp\left( -{1 \over {2\mu^2}} \int~d^2x_T ~\rho^2 (x)\right)
\end{equation}
where $\rho $ is the charge density per unit area.

For such a Gaussian charge distribution we have
\begin{equation}
	<\rho^a(x) \rho^b (y)> = \mu^2 \delta^{ab} \delta^{(2)} (x_T -
y_T)
\end{equation}
where it is straightforward to estimate $\mu^2 = 1.1 A^{1/3}~Fm^2$

The problem we must solve is therefore that described by the theory in
the presence of an arbitrary external source of surface charge on the
light cone and then integrating over all possible values of the
charge.  The fields generated correspond to a stochastic source of
charge.  The problem has therefore been reduced to a many body
theory.  It is possible to integrate out the sources entirely and get
an action in terms of the quark and gluon fields with a term
involving $\mu^2$ which is associated with the valence quark charge
density.  The value of $\mu^2$ sets the scale for the coupling
constant.  For large $\mu^2$ the coupling is small and weak coupling
methods should be reliable

We can now compute the gluon distribution function to lowest order
in $\alpha_s $ and to lowest order in $\mu^2$.  We first compute the
change in the propagator induced by the sources.  This is
\begin{equation}
	\delta < A^\mu A^\nu >  =  = \int [d\rho]~ g^2 \delta^{\mu i}
\delta^{\nu j} < \left( {{\nabla^i_T} \over {\partial^+ \nabla^2_T}}
\right)
\rho^a (x) \left( {{\nabla^i_T} \over {\partial^+ \nabla^2_T}} \right)
\rho^b (x) >
\end{equation}
or
\begin{equation}
\delta D^{\mu \nu} (k,q) = g^2 \mu^2 \delta^{ab} (2\pi)^4 \delta (k^-)
\delta (q^-)  \delta^{(2)} (k_T - q_T ) \delta^{\mu i}
\delta^{\nu j} {{k_T^i q_T^j} \over {k^+ q^+ k_T^2 q_T^2}}
\end{equation}

Using this form of the propagator, it is now easy to show that
\begin{equation}
	{1 \over {\pi R^2}} {{dN} \over {d^3k}}
= {{\alpha_s \mu^2 (N_c^2 -1 )} \over {\pi R^2}} {1 \over
{k^+k_T^2}}
\end{equation}
This is just the Weizsacker-Williams distribution function weighted
by the average charge squared per unit area.  This result reflect the
RMS fluctuations in the stochastic background field induced by the
source associated with the valence quarks.

We will soon see that this result is valid only in the range of
momentum where
$\alpha_s^2 \mu^2 << k_T^2 << \mu^2$  The last term in the previous
limit is just the
region of validity of our derivation of the many body theory.  The
first term is the limit of validity of assuming that $\mu^2 $ is
small and that one can expand to first order in this quantity.

We can generalize our results to all orders in $\alpha_s \mu $ and
first order in $\alpha_s$  To do this, we solve the classical problem
of computing the fields in the background of an arbitrary source and
then integrate over the source.  This classical problem will be
accurate to first order in $\alpha_s$

We must solve the equations of motion
\begin{equation}
      D _\mu F^{\mu \nu} = g\delta^{\nu +} \delta (x^-) \rho (x^+,
x_T)
\end{equation}
There is a solution of these equations of motions with
$A_+ = A_- = 0$ which also has $F_T = 0$.  This solution is of the
form
\begin{equation}
	A_j(x^+x_T) = \theta (x^- ) \alpha_j
\end{equation}
The condition that $F_T = 0$ is equivalent to the condition that the
field $\alpha_j $ is a gauge transformation of the vacuum
configuration of a two dimensional gauge theory.  The condition that
$D_T \cdot E_t = -g\rho $ is equivalent to the two dimensional
gauge condition
\begin{equation}
	\nabla \cdot \alpha = -g\rho
\end{equation}

The above field configuration may be written as
\begin{equation}
	\tau \cdot \alpha = -{1 \over ig} U(x_T) \nabla_i U^\dagger
(x_T)
\end{equation}
and the gauge condition is
\begin{equation}
	\vec{\nabla} ( \cdot U \vec{\nabla} U^\dagger) = ig^2 \rho
\end{equation}

The integration over the sources may be written as
\begin{equation}
	\int [dU] exp \left( - {1 \over {\mu^2 g^4 }} tr \left( \vec{\nabla}_T
\cdot U  {1 \over i} \vec{\nabla}_T U^\dagger \right)^2 \right)
\end{equation}
There is of course a Fadeev-Popov determinant for this measure
which comes from restricting to Feynman gauge, but we will not be
concerned with this measure here.  (The determinant does not affect
the arguments we present for the validity of perturbative
expansions nor the scaling behavior in $k_T$.)

Note that the coupling constant for this theory is $g^2\mu$ so that
the expansion parameter is $g^2\mu /q_T$.  The fluctuations over
the external charge generate an ultraviolet finite theory.  The
correlation function of $U \nabla U^\dagger /g$ generates the
modifications due to the sources of the gluon propagator.
This correlation function should die exponentially at long distances
corresponding to small coupling , that is, in momentum space, the
correlation function should be finite at zero momentum.

The step function $\theta (x^-) $ in the solution of the classical
field equations guarantees that the distribution of gluons is
proportional to $1/k^+$ when the theory is solved to first order in
$\alpha_s$  The gluon distribution function is therefore of the form
\begin{equation}
{1 \over {\pi R^2}} {{dN} \over {dx d^2k_T}} = {{N_c^2-1} \over \pi^2}
{}~{1 \over x}~ {1 \over \alpha_s} H(k_T^2 /\alpha_s^2 \mu^2)
\end{equation}
where $lim_{y \rightarrow \infty } H(y) = 1/y $

\section{Summary}

There are of course many problems which must be solved before this
approach can provide a realistic theory of the distribution functions
at small $x$.  The first order corrections in $\alpha_s$ must
at least be understood.  This will generate the induced
contribution of sea quarks.  It will also presumably lead to the
Lipatov enhancement.  This must be nontrivial since for this theory
where the valence quarks are localized to a delta function on the
light cone, there is no scale of $P^+$, the momentum per nucleon of
the nucleon in the nucleus.  This can only arise from the cutoff
dependence of a regularized delta function.  Such a dependance will
not affect the physics to leading order in $\alpha_s$ nevertheless,
since if we change the cutoff by a finite amount
$(\Lambda /k^+)^{C\alpha_s}$ changes by only an amount proportional
to  $\alpha_s$

In addition to the above problems, the issue of how large a
transverse momentum one can use to define a sensible theory
remains open.  Even though our approximations were only valid
for $q_T << \mu$ the validity of assuming straight line trajectories
along the light cone should be valid in a much broader region.
The extent to which the theory can be solved in the large momentum
transfer region and the extent of the region of validity of the
external source approximation remains open.

Finally, there is the issue of actually computing structure functions
for deep inelastic scattering or Drell-Yan particle production.
We have here only discussed computing the expectation values of
quark distributions in the ground state wavefunction.  When a probe
of momentum $q^2$ is introduced, in general there will be two
parameters of interest $q^2$ and $\mu^2$.  For $q^2 >>
\mu^2$, presumably one can analyze the distribution functions using
Altarelli-Parisi equations.  For $q^2 \le \mu^2$, the situation must
be more complicated since here the momentum of the quarks inside
the hadron wavefunction are important and one is not in the scaling
region.

\section{Acknowledgements}

I thank Stan Brodsky, Miklos Gyulassy,  Al Mueller and Janos Polonyi
for useful comments and insights. This work was stimulated in part by
conversations with Klaus Kinder-Geiger.  I thank Raju Venugopalan
with whom the results described above were derived.    This work
was supported under DOE High Energy DE-AC02-83ER40105 and DOE
Nuclear DE-FG02-87ER-40328.

\end{document}